# ESTIMATING LIMITS FROM POISSON COUNTING DATA USING DEMPSTER–SHAFER ANALYSIS

By Paul T. Edlefsen, Chuanhai Liu and Arthur P. Dempster

*Harvard University, Purdue University and Harvard University*

We present a Dempster–Shafer (DS) approach to estimating limits from Poisson counting data with nuisance parameters. Dempster–Shafer is a statistical framework that generalizes Bayesian statistics. DS calculus augments traditional probability by allowing mass to be distributed over power sets of the event space. This eliminates the Bayesian dependence on prior distributions while allowing the incorporation of prior information when it is available. We use the Poisson Dempster–Shafer model (DSM) to derive a posterior DSM for the "Banff upper limits challenge" three-Poisson model. The results compare favorably with other approaches, demonstrating the utility of the approach. We argue that the reduced dependence on priors afforded by the Dempster–Shafer framework is both practically and theoretically desirable.

**1. Introduction.** This article addresses the problem of estimating a Poisson rate in the presence of additive and multiplicative noise, when additional measurements provide data for estimating the nuisance parameters. The problem of estimating rates from noisy or censored Poisson counting data arises in various subdisciplines of physics. In astrophysics, for example, the counts are photons emitted from a distant source. In particle accelerator experiments, the counts are indirect measurements of the number of particles produced by a high-energy collision. In such contexts the observed counts typically include some additive background noise, such as ambient particles. In many cases there is also multiplicative noise, caused, for example, by photon censoring or particle decay, which further complicates the process of estimating the rate of interest. This article addresses the case in which the physicist hopes to isolate the source rate by taking additional "subsidiary" measurements to estimate the additive and multiplicative noise components.

In particular, we address the case in which three separate counts are observed: one ($y$) for estimating the additive background noise $b$ alone, one ($z$) for estimating the multiplicative noise component $\varepsilon$ alone, and one ($n$) from the main experiment, in which the rate of interest $s$ interacts with the nuisance parameters as $\varepsilon s + b$. We assume that $y$ has a Poisson distribution with rate $tb$, for some known constant $t$, that $z$ has a Poisson distribution with rate $u\varepsilon$, for some known constant $u$, and that $n$ has a Poisson distribution with rate $\varepsilon s + b$. The goal is to provide an estimate of $s$ to some degree of confidence.

We allow for multiple observations of the three counts, for cases in which the nuisance parameters $b$ and $\varepsilon$ differ across measurements. This arises in particle accelerator experiments when the observed counts are of different types of particles: it is often the case that the particle of interest is too short-lived to observe directly, but the particle types to which it decays are measurable. These different particle types are measured separately, with different degrees of noise. We refer to the observation number $i$ as a *channel*, and allow for both $\varepsilon$ and $b$ to vary across channels. In the context of particle accelerator experiments, each multiplicative component $\varepsilon_i$ represents a combination of the rate of decay to particle type $i$ and other factors such as the running time and the accelerator beam intensity.

The model is, for channel $i \in 1, \ldots, N$,

$$
\begin{aligned}
n_i &\sim \mathcal{P}\text{ois}(\varepsilon_i s + b_i), \\
y_i &\sim \mathcal{P}\text{ois}(t_i b_i), \\
z_i &\sim \mathcal{P}\text{ois}(u_i \varepsilon_i).
\end{aligned}
\tag{1.1}
$$

The constants $t_i$ and $u_i$ are given, as well as the observed counts $n_i$, $y_i$, and $z_i$. The parameter of interest is $s$, with $\varepsilon_i$ and $b_i$ considered nuisance parameters. The goal is to estimate and provide confidence limits for $s$.

1.1. *The Higgs particle.* This work is motivated in particular by the problem of estimating the mass of the elusive Higgs particle. The Higgs boson is the only Standard Model (SM) subatomic particle not yet observed. The mass of the particle, if the particle exists, has profound implications for particle physics and for cosmology. Theoretical considerations place the mass somewhere between about 130 and 190 GeV. Previous experimental results suggest that the mass, if the particle exists, is somewhere between 65 GeV to 186 GeV [Igo-Kemenes (2006)]. Masses outside this range are excluded at a specific confidence level by comparing precise measurements with what would be expected from indirect effects if a Higgs boson of a given mass existed. These limits are effectively on the logarithm of the Higgs mass, so by extending the confidence interval a bit, the upper mass limit increases fairly rapidly. If it turns out that the mass is below 130 GeV, then new physics would be required to explain the phenomenon. If the boson does not exist, then the fundamental source of mass in the Universe would not be explained by the Standard Model.

Experiments to determine the mass of the Higgs boson involve complex, expensive equipment. At present the most promising apparatus is the Large Hadron Collider (LHC) at CERN, which hosts several experiments, including ATLAS, a collaboration of over two thousand physicists from 37 countries. The LHC is expected to become fully operational in 2009 and to operate for about a decade. Its total cost will be around 8 billion US dollars. When in operation, about 7000 physicists from 80 countries will have access to the LHC, and the data analysis project is expected to involve many more scientists.

The experiments do not directly observe the Higgs boson, but do detect and measure the particles into which it decays. From these the mass of the Higgs can be calculated, assuming that it is indeed being produced in the observed interactions. Each measured combination of specific particles is called a *channel*, and the fraction of the Higgs particle that decays into each channel is called the *branching ratio*. Physical theory states that the branching ratio is a function of the mass of the decaying particle.

By measuring each channel in the presence of a particle collision with sufficient energy to produce a Higgs particle, and again in the absence of such a collision, the physicists obtain data from which limits on the production rate of a possible Higgs boson of mass $m_H$ can be deduced. By comparing this with the expected rate of Higgs production according to theory, this can be converted into information on excluded or allowed ranges of $m_H$.

Unfortunately the choice of the correct model is not entirely clear, nor is the correct statistical methodology clear, given a particular model. In fact, the data are processed and filtered substantially during the detection process, further complicating the analysis.

Given the importance of the science and the cost of the data acquisition process, any determination of the mass of the Higgs particle is sure to be debated and its analysis methods scrutinized. Groups of scientists have already begun to explore, through simulation studies and theoretical arguments, the confidence that might justifiably be attributed to any future conclusion.

1.2. *The BIRS A1 limits project.* In July 2006 a workshop on statistical inference problems in high energy physics and astronomy was held at the Banff International Research Station (BIRS), bringing together physicists, astronomers, and statisticians to "bring the latest methods to the attention of the scientific community, and to develop statistical theory further by considering special aspects that arise in these scientific contexts" [Linnemann, Lyons and Reid (2006)]. One of the primary objectives of the workshop was to address issues of statistical significance in the presence of nuisance parameters such as those that arise in the determination of the mass of the Higgs particle. By the end of the workshop the participants decided that the various methods for determining confidence limits on the Higgs mass should be allowed to compete in a simulated experiment [Linnemann, Lyons and Reid (2007)]. This open challenge was called the BIRS A1 Limits Project [Heinrich (2006a)].

Data were provided as counts from the model in equation (1.1). As further described below, single-channel and 10-channel data were provided in a total of 3 tasks. Respondents to the challenge submitted 90% and 99% upper (one-sided) confidence bounds on $s$, given the 3 counts per channel.

Previous work on this model include a Bayesian approach for one channel in Heinrich et al. (2004) and Demortier (2005) and for multiple channels in Heinrich (2005), a Profile Likelihood approach in Rolke, Lopez and Conrad (2005), a frequentist-Bayesian hybrid approach in Conrad and Tegenfeldt (2006), and fully frequentist approaches in Punzi (2005) and Cranmer (2003). Prior to this challenge, these approaches had never been compared using the same datasets or criteria. No attempt had been made previously to use a hierarchical Bayesian approach, or to use Dempster–Shafer analysis. As a result of the challenge, all of these approaches may now be compared on common ground.

1.3. *Dempster–Shafer.* We submitted upper limits to the Banff challenge based on the Dempster–Shafer (DS) method described below. Our method uses DS analysis [Dempster (1968a); Shafer (1976); Dempster (2008)] to construct a Bayesian-style posterior using no priors. Dempster–Shafer analysis utilizes an extended probability calculus that assigns probability mass to elements of the powerset of a state space rather than to elements of the state space itself. In the context of real-valued parameters, DS models are maps from *ranges* (and sets of ranges) of the parameter space to real numbers, rather than maps from single parameter values to real numbers as is the typical case in frequentist and Bayesian calculi. As such, these standard approaches are subsets of the DS calculus.

In the 40 years since its introduction, aspects of the Dempster–Shafer theory have become widely used in several disciplines, notably in operations research, fuzzy logic, and Bayesian networks. Unfortunately its acceptance in mainstream statistics has been hindered by inconsistent presentation. As in many realms of statistics, there is no standard language for describing DS, and its proponents often disagree on emphasis and even philosophy. The moniker "belief function," commonly applied to DS models, has contributed to a suspicion that the theory is nonstatistical or overly subjective. Rather than dismiss DS as ungrounded, we consider its foundations firm and seek to investigate its potential to contribute to applications such as this one.

We encourage the skeptical reader to consider the results and the theory described in this paper on their own merit.

The remainder of this article describes results from the Banff challenge and from our own simulation studies, then provides a high-level description of the DS approach to this problem, accompanied by an introduction to the general framework of DS analysis. Additional results and derivations of the mathematics are provided in an Appendix and as Supplementary Materials [Edlefsen, Liu and Dempster (2008)].

## 2. Results.

2.1. *The Banff challenge.* The official problem statement for the Banff challenge may be found in Heinrich (2006b) and on the BIRS A1 Limits Project website [Heinrich (2006a)]. Three datasets were generated by Joel Heinrich, each corresponding to a separate task of the challenge [Heinrich (2006a)]. The tasks are numbered 1a, 1b, and 2.

The data provided for task 1a consist of 60,229 independent sets of single channel ($N = 1$) counts $(n, y, z)$. In all cases, $t = 33$ and $u = 100$. The data provided for task 1b consist of 39,700 permutations of single-channel counts for $n = 0, \ldots, 49$, $z = 0, \ldots, 30$, and $y = y_l, \ldots, y_u$, for $y_l$ varying from 0 to 7, and $y_u$ varying from 13 to 24, depending on $z$. In all cases, $t = 3.3$ and $u = 10$. The data provided for task 2 consist of 70,000 independent sets of ten channel ($N = 10$) counts $(n_1, x_1, y_1, \ldots, n_{10}, x_{10}, y_{10})$. $t_i = 15 + 2i$ and $u_i = 53 + 2i$ $\forall i \in 1, \ldots, 10$.

Subsequent sections of this article outline and justify the Dempster–Shafer approach that we used to generate our submissions to the Banff challenge. We applied our DS method to the three datasets corresponding to tasks 1a, 1b, and 2. We implemented the algorithm in the Perl programming language, using a rectangle integration procedure to numerically approximate the integrals in Supplementary Materials equations (S.5) and (S.6) for each of 100 values of $x$. The 100 values of $x$ were chosen separately for each data channel, with the maximum value chosen to include the nonnegligible range of both $F_{\mathbf{S}_l^i}(x)$ and $F_{\mathbf{S}_u^i}(x)$ for that channel's values of $(n, y, z, t, u)$. After combining the values of $(F_{\mathbf{S}_l^i}(x) - F_{\mathbf{S}_u^i}(x))$ from each channel $i$, we reported the 90th and 99th percentiles of $f_{\mathbf{S}}(x)$ as the 90% and 99% one-sided upper bounds on $s$.

2.2. *Evaluation.* The question of the appropriate evaluation procedure is discussed at length in Heinrich et al. (2004). The primary metric used for the Banff challenge was the actual coverage, discussed there among several other possibilities. A secondary metric, Bayesian credibility, was also computed for each submission. An obvious additional metric of interest is the length of the submitted intervals, since it would be possible with 0-length and infinite-length intervals to achieve any desired coverage, but these would be meaningless. Our submissions show desirable traits in all three categories, and compare favorably to the other submissions and to previously published results.

2.2.1. *Coverage.* All three tasks were evaluated by calculating or estimating the actual coverage $C(s)$ of the submitted 90% and 99% intervals. Coverage is the proportion of intervals that cover the true value. For the single-channel case, if we could generate all possible sets of triples $(n, y, z)$, and if we computed the corresponding confidence limits $R(n, y, z)$ for each set, then the coverage could be calculated as the fraction of the intervals that cover the true value of $s$, each weighted by the probablity of generating that set:

$$(2.1) \quad C(s) = \sum_{(n,y,z) \text{ s.t. } s < R(n,y,z)} e^{-\mu} \frac{\mu^n}{n!} e^{-\nu} \frac{\nu^y}{y!} e^{-\rho} \frac{\rho^z}{z!},$$

where

$$\mu = \varepsilon s + b,$$
$$\nu = tb,$$
$$\rho = u\varepsilon,$$

and $R(n, y, z)$ is the upper bound result submitted by the contender.

Of course, since there is an infinite number of possible sets, a Monte Carlo approximation has to suffice. The most obvious approximation is to sample the intervals from their joint distribution and compute the observed fraction covering the true value of $s$. The actual method used by Joel Heinrich improved upon this by using importance sampling to allow each interval to contribute to coverage calculations at multiple values of $s$.

The 101 values of $s$ used to evaluate the three tasks, known only to the evaluator until all submissions were received, were 0–25 in increments of 0.25. The values of the nuisance parameters for the coverage calculation (revealed after the challenge) were $\varepsilon = 1$, $b = 3$ for tasks 1a and 2, and $\varepsilon = 0.1$ and $b = 0.3$ for task 1b. These task 1b values were chosen to enable a complete enumeration of data sets, for which $C(s)$ would be nonnegligible (for all of the 101 $s$ values), to calculate coverage with improved precision.

The coverage results for task 1a are shown in Figure 1. The results for task 1b are shown in Figure 2. The results for task 2 are shown in Figure 3. The grey line on each plot indicates the desired coverage. Note that the 90% plots are on a different scale than the 99% plots. For tasks 1a and 2, error bars indicate two standard errors (from the importance sampling described above) on either side of the coverage estimate. Task 1b errors are neglible.

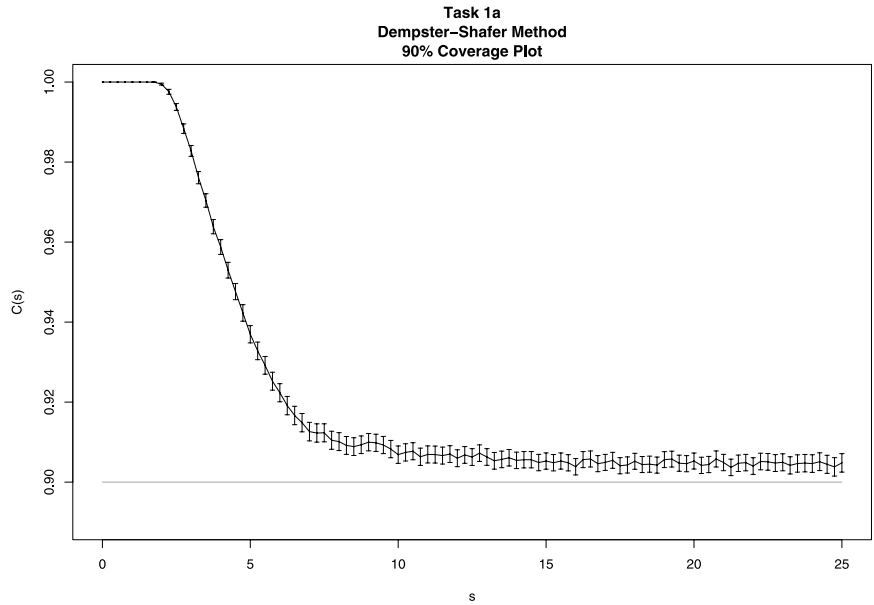

(a) 90% Coverage

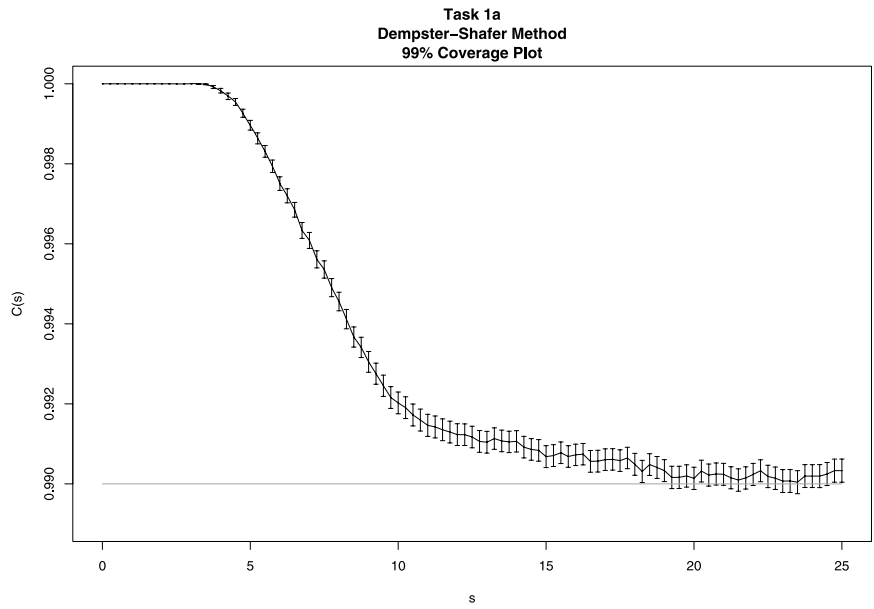

(b) 99% Coverage

FIG. 1. *Task 1a coverage plots. This is the "moderate uncertainty" single-channel task. The plots show estimated coverage and 95% CI. The target coverage is shown in grey. For sufficiently large s values, the DS method produces coverage results close to the target.*

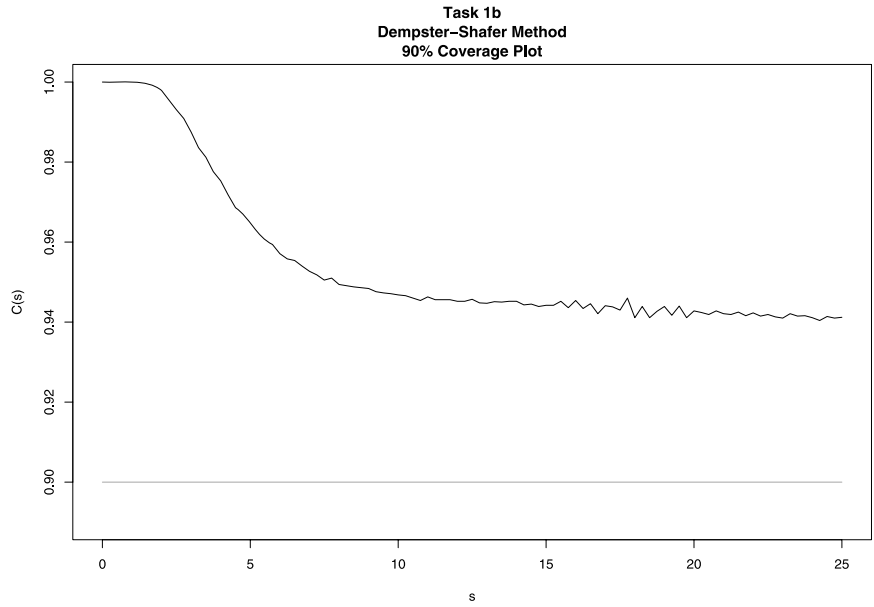

(a) 90% Coverage

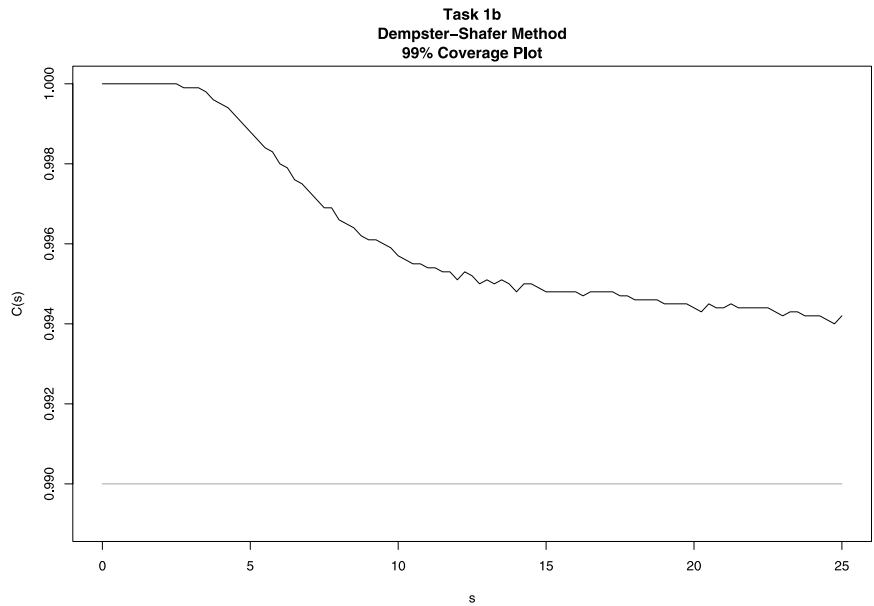

(b) 99% Coverage

FIG. 2. *Task 1b coverage plots. This is the "large uncertainty" single-channel task. For these data the coverage is computable with negligible error. The target coverage is shown in grey. The DS method slightly overcovers on this task.*

ESTIMATING POISSON LIMITS USING DSA 9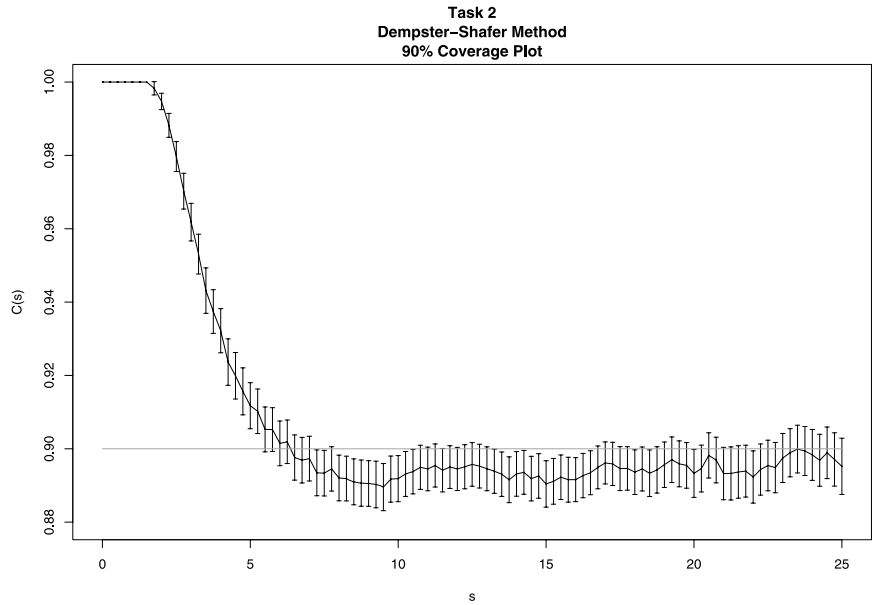

(a) 90% Coverage

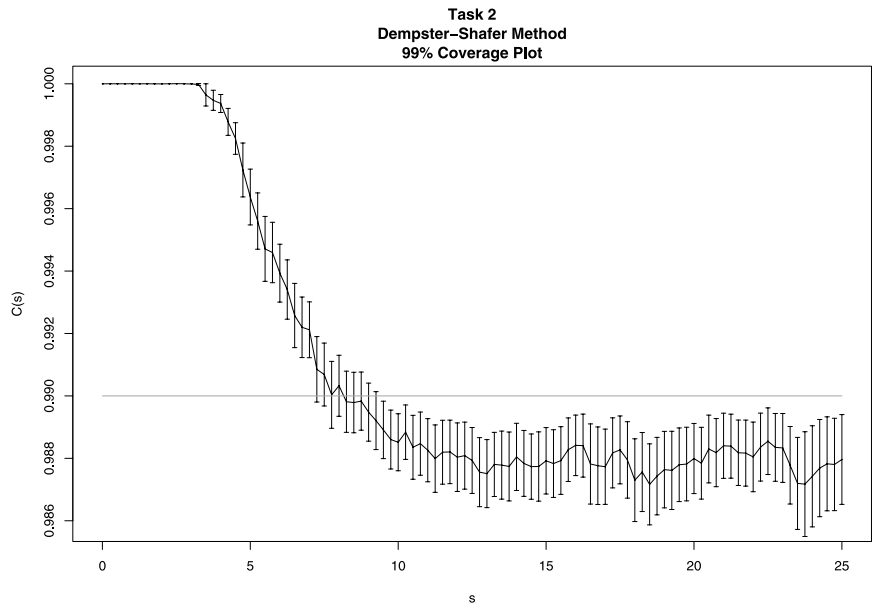

(b) 99% Coverage

FIG. 3. *Task 2 coverage plots. In this task there are ten channels. The plots show estimated coverage and 95% CI. The target coverage is shown in grey. For sufficiently large s values, the DS method produces coverage results close to the target.*

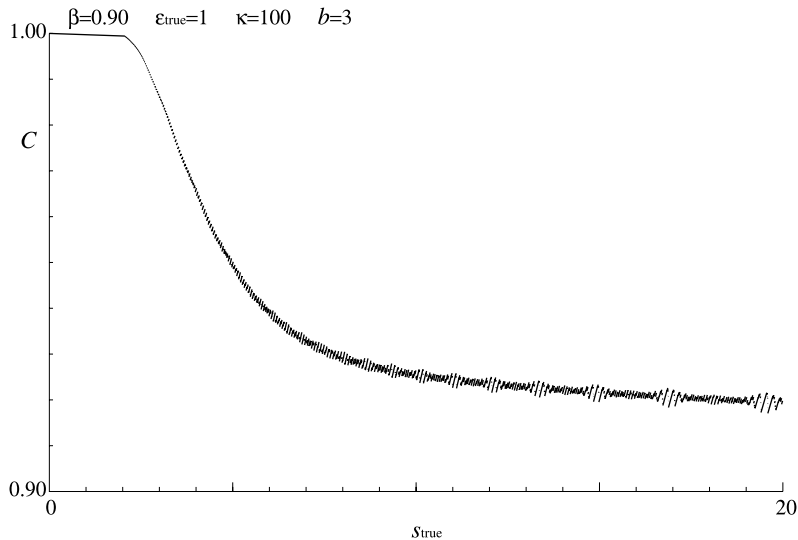

FIG. 4. *Pure Bayesian approach (with known b) 90% coverage. This figure [from Heinrich et al. (2004)] was produced using a Bayesian procedure for the somewhat simpler single-channel case in which b is fixed and known. These results exhibit the same overcoverage phenomenon at low s values as that seen in the DS results.*

The coverage results for tasks 1a and 2 were very close to the desired coverage after an initial period of overcoverage. The initial period of overcoverage, also found in Heinrich et al. (2004), is inevitable at $s = 0$ and understandable at small values of $s$, because the upper limits will tend to lie above the true small $s$ values there. The results for task 1b show actual coverage closer to 95% and 99.5%, slightly above the desired values of 90% and 99%. This is presumably due to the smaller values of $t$ and $u$ used in this task. In the project specification, task 1b is referred to as the "large uncertainty" task because the nuisance parameters are more difficult to estimate.

Overall, the results are impressively accurate. For comparison, we include a figure from Heinrich et al. (2004) showing the result of applying a pure Bayesian approach to a similar problem (but in which the background $b$ is fixed and known rather than a nuisance parameter). Note that in this figure (Figure 4) $\kappa$ is what we refer to as $u$. At $\kappa = 100$, it corresponds to the value used in task 1a.

2.2.2. *Credibility.* The submissions were also evaluated by calculating their Bayesian credibility, which is the probability of the submitted limit exceeding a value of $s$ drawn from its marginal posterior distribution, for some particular model and priors. If the model and priors used for this calculation were the same as those used to generate the limits, then, by definition, the credibility for the 90% limits would be 90%, and for the 99% limits, 99%. The actual model used to calculate the credibility puts a flat prior on $s \geq 0$, and independent gamma prior distributions on each $b$ and $\varepsilon$. For tasks 1a and 2 the gamma prior on $b$ has shape 300 and scale 0.01, so that its mean is 3 and its standard deviation is 0.3, and the gamma prior on $\varepsilon$ has mean 1 and standard deviation 0.1. For task 1b these priors have means 0.31 and 0.1, respectively (matching the values used for the coverage calculation), and standard deviations of 0.1 and 0.03, respectively. The software used to calculate the credibility is described in Heinrich (2005).

2.2.3. *Length.* As discussed previously, it would be possible to achieve perfect coverage by submitting a certain fraction of infinite-length intervals, with the rest being zero-length. None of the contenders did this, as it clearly violates the spirit of the competition. Since $s$ is nonnegative, the lengths of the submitted upper limits are simply the limits themselves. We calculated the quantiles of the 90% and 99% upper limits for each of the submitted methods, and found that there is little discernable difference among them (excluding those methods with very poor coverage).

The DS method compared favorably with the other submissions. A brief summary of the Banff challenge results is published in Heinrich (2008), and a more detailed comparison is available from the challenge organizers by request. The nonBayesian methods (other than ours) suffered from credibility aberrations, though were generally easier to compute than the full Bayesian methods. Our method had consistent credibility, near target coverage, and was simple and efficient to compute (it does not require sophisticated posterior sampling techniques). Our method is also the most flexible with respect to priors: the DS approach enables the use of priors on the nuisance parameters if prior information is available, but does not require it.

2.3. *Simulation study.* We also compared the DS method to a simple Bayesian alternative by drawing datasets from the model at fixed values of $s$, $b$, and $\varepsilon$ (for one channel). We applied the methods and computed coverage, length, and credibility for each one. We used a straightforward Bayesian procedure that puts a conjugate gamma prior on $tb$, another independently on $u\varepsilon$, and a third independently on $\varepsilon s + b$. This method, described in detail below, requires the same calculations as those used to compute the DS solution, so we performed the comparison using the same implementation of the relevant functions. Since the results depend on the choice of prior, we ran the Bayesian method multiple times with different priors.

For each of the 161 values of $s$ in the range 0–40 (in increments of 0.25), we generated 10,000 sets of $(n, y, z)$ values from the model in equation (1.1). We used the same values of $t$ and $u$ as those used in task 1a of the Banff challenge (33 and 100), and simulated with the same values of the nuisance parameters as were used to evaluate task 1a ($b = 3$, $\varepsilon = 1$). We calculated the Bayesian one-sided (upper) posterior interval four times per dataset, with the following priors: *B1*, which puts independent unit-scale $\mathcal{G}$amma(1)s on each of the three rates *tb*, *u*$\varepsilon$, and ($\varepsilon s + b$); *B2*, which puts a $\mathcal{G}$amma(2) on each; *upper*, which puts a $\mathcal{G}$amma(2) on ($\varepsilon s + b$) and a $\mathcal{G}$amma(1) on the others; and *lower*, which puts a $\mathcal{G}$amma(1) on ($\varepsilon s + b$) and a $\mathcal{G}$amma(2) on the others. A subset of the results are summarized in Table 1. These results are based on the *s* values in the range 20–40, since (for all methods) the range below 20 exhibits the initial overcoverage described above. The table shows averages and standard deviations over the coverages at each *s* value. Medians, not shown, are in all cases very close to the given means. None of these methods displayed credibility or length aberrations.

Of these choices for priors for the Bayesian procedure, the *upper* and *B2* priors produced the best results. In both cases, the average coverages are within two standard deviations of the target coverage. The simulation results for the *DS* procedure show average coverage within one standard deviation of each target.

These results also demonstrate that the *lower* procedure produces intervals that are shorter than those produced by the *upper* procedure. In the next section we will explain that our DS procedure is in some sense a compromise between the *upper* and *lower* prior choices, but these results demonstrate that the DS procedure produces intervals that are preferable to those produced by any of these other methods.

**3. The join tree theorem.** The Bayesian approach used to calculate credibilities is presented in two publications by the challenge organizers [Heinrich et al. (2004); Heinrich (2005)]. That approach and the approach that we used in our simulation study both implicitly apply the *join tree* theorem due to Shenoy and Shafer (1986) or Kong (1986), which is a fundamental tool of Dempster–Shafer calculus. In this section we elaborate on our Bayesian approach and make explicit its use of the join tree (or *junction tree*, or *Markov tree*) theorem. In the next section we build on this foundation to construct the Dempster–Shafer solution.

TABLE 1
*Coverage results from the simulation study. 90% and 99% mean coverages are given for each of the four prior choices with the Bayesian procedure, and for the DS approach. Standard deviations are also given, which reflect the coverage variability across simulation runs. The B2, upper, and DS averages are all within two standard deviations of the target coverage*

| Interval (stat) | B1     | B2     | lower  | upper  | DS     |
|-----------------|--------|--------|--------|--------|--------|
| 90% (mean)      | 0.8720 | 0.8953 | 0.8616 | 0.9038 | 0.9032 |
| 90% (stdev)     | 0.0051 | 0.0034 | 0.0047 | 0.0033 | 0.0034 |
| 99% (mean)      | 0.9855 | 0.9898 | 0.9838 | 0.9910 | 0.9901 |
| 99% (stdev)    | 0.0017 | 0.0011 | 0.0017 | 0.0011 | 0.0011 |

The authors of Heinrich et al. (2004) describe the posterior distribution of $s$ for the single-channel tasks in the case of fixed $\varepsilon$ and $b$, and then extend this to the case when "subsidiary measurements" yield gamma distributions for these nuisance parameters. With conjugate gamma priors for $u\varepsilon$ and $tb$, the observations of $y$ and $z$ yield gamma-distributed posteriors. These posteriors are then incorporated as priors on $b$ and $\varepsilon$ in the relation $n \sim \mathcal{P}\text{ois}(\varepsilon s + b)$.

This approach is an example of the *belief propagation* algorithm for join trees [Pearl (1982); Shenoy and Shafer (1986); Kong (1986)]. The posteriors for $b$ and $\varepsilon$ are calculated given $y$ and $z$ (and $t$ and $u$), then this information is used in computing a posterior for $s$, given $n$. This is mathematically equivalent to computing the joint distribution of $(s, b, \varepsilon)$ given $(n, y, z, t, u)$ and then marginalizing, but is more convenient. In general, belief propagation can be far more efficient than marginalizing the complete joint posterior.

The Bayesian approaches of Heinrich et al. (2004) and Heinrich (2005) put a prior directly on the quantity of interest, $s$. The Bayesian approach that we used for the simulation study differs only in that we put a prior instead on the quantity $(\varepsilon s + b)$, with an additional indicator that $s \geq 0$. Any proper joint prior on $(s, \varepsilon, b)$ can be converted to a joint prior on $((\varepsilon s + b), \varepsilon, b)$, and vice-versa so long as we ensure that the support of $s$ remains nonnegative, so in this context the choice is simply a matter of convenience.

The Bayesian and DS approaches consider the quantities of the model in equation (1.1) as random variables, except for the constants $t$ and $u$. We are interested in the posterior distribution of $\mathbf{S}$, which is a margin of the joint posterior of $(\mathbf{S}, \mathbf{E}, \mathbf{B})$ given $(\mathbf{N}, \mathbf{Y}, \mathbf{Z})$ and $(t, u)$. If we define an auxiliary random variable $\mathbf{L}_n := \mathbf{ES} + \mathbf{B}$, then the Banff model gives us $(\mathbf{N}|\mathbf{L}_n = l_n) \sim \mathcal{P}\text{ois}(l_n)$, and the posterior distribution of $\mathbf{L}_n$, given $\mathbf{N} = n$, has a gamma distribution (so long as we choose a conjugate prior for $\mathbf{L}_n$).

Using the join tree theorem, we can first compute the posteriors of $\mathbf{L}_n$, $\mathbf{B}$, and $\mathbf{E}$, given the observed data, and then use these distributions to compute the posterior distribution of $\mathbf{S} = \frac{\mathbf{L}_n - \mathbf{B}}{\mathbf{E}}$. Suppose that the "subsidiary measurements" yield gamma distributions on $\mathbf{L}_n$, $\mathbf{B}$, and $\mathbf{E}$, with shape parameters $k_n$, $k_b$, and $k_e$, respectively, and scale parameters $w_n$, $w_b$, and $w_e$, respectively. Then $\mathbf{S}$ has the distribution of $\mathbf{S}^*|(\mathbf{S}^* \geq 0)$, where

$$\mathbf{S}^* \sim \frac{\mathcal{G}\text{amma}(k_n, w_n) - \mathcal{G}\text{amma}(k_b, w_b)}{\mathcal{G}\text{amma}(k_e, w_b)}.$$

That is,

(3.1)
$$F_\mathbf{S}^*(x) = \mathbb{P}\left(\frac{\mathbf{L}_n - \mathbf{B}}{\mathbf{E}} \leq x\right) \quad \text{and}$$

$$F_\mathbf{S}(x) = \frac{F_\mathbf{S}^*(x) - \mathbb{P}(\mathbf{S} < 0)}{1 - \mathbb{P}(\mathbf{S} < 0)}.$$

Noting that $\mathbf{S} < 0$ whenever $\mathbf{L}_n < \mathbf{B}$, we get

$$\text{F}_{\mathbf{S}}(x) = \frac{\mathbb{P}(\mathbf{L}_n \leq \mathbf{B} + x\mathbf{E}) - \mathbb{P}(\mathbf{L}_n < \mathbf{B})}{1 - \mathbb{P}(\mathbf{L}_n < \mathbf{B})}, \tag{3.2}$$

which may be expressed in terms of the $\mathcal{B}$eta CDF (after some manipulation; see the Appendix for a complete derivation) as

$$\begin{aligned}
\text{F}_{\mathbf{S}}(x) = 1 - &\bigg( p\mathcal{B}(\alpha, k_e, k_n) \\
&- \int_0^\alpha p\mathcal{B}\bigg(\frac{w_b}{w_b + (1 - \gamma/\alpha)}, k_e + k_n, k_b\bigg) d\mathcal{B}(\gamma, k_e, k_n) \, d\gamma \bigg) \\
&\times \bigg( p\mathcal{B}\bigg(\frac{w_n}{w_n + w_b}, k_b, k_n\bigg) \bigg)^{-1},
\end{aligned} \tag{3.3}$$

where $d\mathcal{B}(\cdot, a, b)$ is the PDF of a $\mathcal{B}$eta distribution with parameters $a$ and $b$, $p\mathcal{B}(\cdot, a, b)$ is its CDF, and $\alpha := \frac{w_n}{w_n + xw_e}$.

We used equation (3.3) to compute the Bayesian methods that we compared to our DS method in the simulation study described above. Our choice of priors determined the shape parameters $k_n$, $k_b$, and $k_e$, while the scale parameters $w_b$ and $w_e$ were the inverse of $t$ and $u$, respectively.

Although in general the results of a Dempster–Shafer analysis are not necessarily expressible in such simple terms, in this case our DS approach is effectively a compromise between the prior choices we labeled *upper* and *lower*. As we will explain below, the DS framework represents uncertainty in terms of mass distributions over ranges of the parameter space, and in this instance the approach yields ranges bounded by random variables whose marginal CDFs are the same as the CDFs of the Bayesian posteriors on $\mathbf{S}$ resulting from the *upper* and *lower* prior sets. The DS approach in this case effectively treats the posterior pdf of $\mathbf{S}$ as proportional to the difference between those Bayesian posterior CDFs.

**4. Dempster–Shafer theory.** The Dempster–Shafer theory extends traditional Bayesian/frequentist statistics by appending a third category "don't know" to the familiar dichotomy "it's true"/"it's false." The theory assigns to any assertion a probability $p$ *for* that assertion, a probability $q$ *against* that assertion, and a third probability $r = 1 - p - q$ that remains effectively unassigned. Although the actual state of the described phenomenon is understood to be constrained such that the assertion is either true or false, DS theory allows you, the observer, to describe your evidence for and against the assertion without the traditional constraint that all such evidence be construed unambiguously. The remaining probability, $r$, represents your residual uncertainty after assessing the available evidence.

A more mathematical explanation of DS theory may help to clarify the high-level description just given. The mathematical framework of DS theory is that of random sets. From this perspective, if $p$ is the probability of the event $T$ (that the assertion is true), and $q$ is the probability of the event $F$ (that the assertion is false), then $r$ is the probability of the set $\{T, F\}$ ("don't know"). The DS mass function $m(A): 2^S \mapsto [0, 1]$ is mathematically indistinguishable from a probability measure over an extended state space (the power set of the event space $S$). DS theory combines the logic of set theory with this random sets framework to yield a powerful calculus for reasoning about uncertainty.

If we define an assertion $A \subset S$ as a set of events (in words, "the true state of the described phenomenon is in the set $A$"), then the accumulated evidence for that assertion is given by $p(A) = \sum_{B \subseteq A} m(B)$. That is, it includes evidence that is unambiguous ("the true state is $e$, an element of $A$") and evidence that is ambiguous ("the true state is in the set $A'$, a subset of $A$ with cardinality greater than 1"). The total evidence against the assertion $A$ is given by $q(A) = p(A^c) = \sum_{C: A \cap C = \varnothing} m(C)$. Any evidence that is ambiguous with respect to the assertion $A$ is accumulated in the residual uncertainty $r(A) = 1 - p(A) - q(A)$, which is the sum of the evidence on sets that overlap both $A$ and $A^c$.

When all evidence is unambiguous, DS theory coincides completely with Bayesian (and frequentist) statistics, with $p(A) = \sum_{e \in A} \mathbb{P}(e) = 1 - q(A)$. What the DS framework adds is the ability to tolerate ambiguous evidence, which is particularly useful when describing the joint distribution of non-independent margins: with the Bayesian constraint that $r(\cdot) = 0$, marginal evidence must be combined with evidence or assumptions about conditional distributions when extending that evidence to a joint state space. The DS framework allows the observer to remain agnostic when extending evidence on margins into a joint space. For instance, if two nonindependent Bernoullis (with respective marginal probabilities $p_1$ and $p_2$ of states $T_1$ and $T_2$) are described by a joint Dempster–Shafer model (DSM), the evidence $p_1$ on $T_1$ in the first margin is, in the joint model, construed as evidence on the set $\{(T_1, F_2), (T_1, T_2)\}$. The Bayesian constraint would require the ambiguity between the constituent states $(T_1, F_2)$ and $(T_1, T_2)$ to be resolved before analysis could proceed.

This feature of the DS framework can be exploited by Bayesians wishing to minimize dependence on convenience priors. Historically, post-analysis depictions of uncertainty about model parameters of interest have been restricted to point estimates and confidence regions. Bayesian posterior distributions are richer depictions that can be summarized with point estimates and credibility regions as needed. DS models are richer still, and just as full Bayesian posteriors have gained acceptance with the passage of time and the improvement of mathematical and computational tools to store them and compute with them, we predict that full DSMs will ultimately gain acceptance as intermediate and final products of statistical analysis. In the interim it is nevertheless convenient for Bayesians who wish to report standard posterior distributions, or for frequentists who wish to report confidence regions, to combine the prior-free analysis summarized by a DSM with a nonambiguity constraint, as we do for the Banff challenge. This approach provides all of the benefits of the Bayesian paradigm with a great deal of additional flexibility and a reduced dependence on priors.

The principal operations of the Dempster–Shafer calculus (DSC) involve extending marginal evidence to a joint space, combining the evidence in the joint space, and projecting from a joint space to a margin. Evidence is represented using DSMs, which, as previously discussed, are essentially probability measures over power sets of state space models (SSMs). In our previous example of two Bernoulli SSMs, the joint DSM was created by first extending the Bernoulli distributions on the two margins to the joint state space (yielding two DSMs with mass functions mapping from the power set of the 4-element joint SSM). Assuming that the evidence yielding the values $p_1$ and $p_2$ can be considered mutually noncompromising (that is, our evidence that the probability of $T_1$ is $p_1$ is independent of our evidence that the probability of $T_2$ is $p_2$), combination is a straightforward multiplication operation, generalizing the Bayesian operation of multiplying likelihoods.

The combination operation is always performed with two DSMs over the same SSM (in our Bernoulli case, we first extend the marginal DSMs to the joint SSM, then combine). If $\mathbf{C}$ is the DSM that is the result of combining DSMs $\mathbf{D}_1$ and $\mathbf{D}_2$ (written $\mathbf{C} = \mathbf{D}_1 \oplus \mathbf{D}_2$), then the combined evidence $m_{\mathbf{C}}(A)$ on any set $A$ is given by accumulating relevant portions of evidence from the two constituent DSMs:

$$m_{\mathbf{C}}(A) = \sum_{A_1, A_2 : A_1 \cap A_2 = A} m_{\mathbf{D}_1}(A_1) m_{\mathbf{D}_2}(A_2).$$

This operation is most conveniently performed using the *commonality* set function $c(A) = \sum_{B \supseteq A} m(B)$, since $c_{\mathbf{D}_1 \oplus \mathbf{D}_2}(A) = c_{\mathbf{D}_1}(A) c_{\mathbf{D}_2}(A)$ [Thoma (1989, 1991); Kennes (1992)]. Thus, the DS combination operation is simply computed by multiplying commonalities of like sets, generalizing the Bayesian combination operation of multiplication of probability masses (or mass densities) on like elements of the SSM. As with the corresponding Bayesian computation, the resulting DSM will usually be normalized to remove mass on the empty set, although normalization may be postponed to the end of the analysis, or avoided altogether if proportional values are sufficient.

A useful conceptualization of Dempster–Shafer models is that of the multivalued map. Dempster's earliest work on the subject (1967a, 1967b, 1968a, 1968b) described the theory in this way, and more recent work by Kohlas,

Monney, and others [see Kohlas and Monney (1994) for a review] has emphasized this perspective, in which a DSM's mass function $m(\cdot)$ is derived from an associated probability distribution $\mathbb{P}$ over an auxiliary state space $\Omega$ and a function $\Lambda : \Omega \mapsto 2^S$. Then $m(A) = \sum_{\omega \in \Omega : \Lambda(\omega) = A} \mathbb{P}(\omega)$. In what follows we will use the prefix "a-" when referring to the associated probability model, or to any auxiliary probability distribution that is used to characterize a DSM.

A simple example that illustrates DS analysis is the case of repeated flips of a single bent coin (with an unknown probability $p$ of landing heads-up). The Binomial DSM for $n$ trials is a joint DS model over the state space of $k \in \{0, \ldots, n\}$ and $p \in [0,1]$. We may summarize our evidence about $p$ as the projection of this joint DSM onto the $p$ margin. We can condition on the observation of $k$ heads by first combining the joint DSM with a deterministic one, effectively placing 0 mass on any set in the joint space that contradicts the observed number of heads. An obvious associated probability model is the Uniform distribution: if $X \sim \mathcal{U}(0,1)$, then the indicator that $X \leq p$ has a Bernoulli distribution, so if we define $\Lambda_h(x) = \{p' : p' \geq x\}$, then the tuple $(\Omega = [0,1], \mathbb{P} = f_{\mathcal{U}}(\cdot), \Lambda = \Lambda_h, S = [0,1])$ defines a DSM over the state space $S$ of possible values of $p$ when we have observed a coin flip and it is heads [if it were tails, we would use $\Lambda_t(x) = \{p' : p' < x\}$]. Such a tuple, called a "hint" by Kohlas and Monney (1995), summarizes our evidence about the unknown parameters of interest implied by an observation. Combining $n$ such hints yields the projection of the conditioned Binomial DSM onto the $p$ margin. This could be equivalently expressed as a single hint with an $n$-dimensional Uniform associated probability model ($x_i \sim \mathcal{U}(0,1)$ for $i \in 1, \ldots, n$) and a multivalued map $\Lambda(x_1, \ldots, x_n) = \{p' : p' \geq x_i \Rightarrow i \in H\}$ (where $H$ is the index set of coins that came up heads). In words, if we observe that $k$ of $n$ coin flips are heads, then $p$ is somewhere between the $k$th and $(k+1)$th ordered Uniforms. That is, for any interval $A := (l, u)$, the mass function $m(A)$ is proportional to the joint density $f(l, u)$ of the $k$ and $k+1$ order statistics of $n$ independent Uniforms.

Note that the Binomial DSM describes $p$ as contained in an a-random interval $(L, U)$ ($L \leq p \leq U$), where $L \sim \mathcal{B}\text{eta}(k, n-k+1)$ and given $L$, $\frac{U-L}{1-L} \sim \mathcal{B}\text{eta}(1, n-k)$. The *a-random* quantities $L$ and $U$ do not describe the distribution of $p$ in the usual sense. In order to obtain a distribution (a *precise* DSM) for $p$, all of the mass in the DSM must be restricted to the singleton sets. Combining any DSM with any precise DSM will yield a precise DSM. The Bayesian practice of combining a likelihood with a prior distribution, for instance, ensures that the resulting posterior is precise. Combining the Binomial DSM for $p$ with a Bayesian prior leads to the same result as the corresponding Bayesian analysis. A Uniform prior, for instance, yields a $\mathcal{B}\text{eta}(k+1, n-k+1)$ posterior.

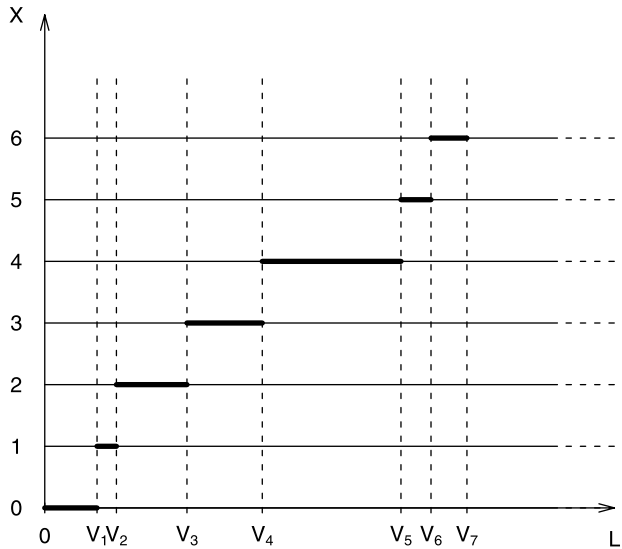

FIG. 5. *The SSM for the Poisson DSM[1] consists of lines in $(X, L)$ space. A typical a-random subset is the union of intervals at levels $X = 0, 1, 2, \ldots$, where each interval corresponds to the waiting time of an auxiliary Poisson process before it transitions out of the state given by the level $X$.*

Combination with an uninformative (uniform) prior is an example of the *plausibility transform* [Cobb and Shenoy (2006)], a method for transforming between a DSM and a probability distribution, which takes $\mathbb{P}(e) \propto c(\{e\})$ for all elements $e \in S$. Glenn Shafer coined the term "plausibility" to refer to the total evidence not in contradiction with an assertion: $\text{Plaus}(A) = 1 - q(A) = 1 - p(A^c) = p(A) + r(A)$. For singleton sets, $\text{Plaus}(\{e\})$ is just the commonality $c(\{e\})$. In the Binomial DSM example, the plausibility transform yields $\mathbb{P}(e) \propto \mathbb{P}_{L,U}(e \in [L, U])$, which turns out to be the $\mathcal{B}\text{eta}(k+1, n-k+1)$ density.

4.1. *The Poisson DSM.* Our solution to the Banff challenge uses the Poisson DSM, which is the Poisson analogue to (and limit of) the Binomial DSM. The full state space model of the Poisson DSM is the cross of the natural numbers $\mathcal{N}$ (for the count, $X$) with the nonnegative Reals $\mathcal{R}^{0+}$ (for the rate, $L$). Figure 5 depicts the full SSM. Conditioning on $L = \lambda$, the $X$ margin has a $\mathcal{P}\text{ois}(\lambda)$ distribution (a precise DSM). For inference about $L$, we are concerned with the DSM on the $L$ margin after conditioning on an observed count $X = k$.

The Poisson DSM is defined mathematically by assigning a mass distribution over a-random subsets of its $(L, X)$ state space. These subsets are determined by an auxiliary sequence of a-random points $0 \leq V_1 \leq V_2 \leq V_3 \leq \cdots$ on the $L$ axis. As illustrated in Figure 5, the auxiliary sequence $V_1, V_2, V_3, \ldots$ defines a corresponding sequence of intervals $0 \leq V_1$, $V_1 \leq V_2$, $V_2 \leq V_3$, $\ldots$ at respective levels $X = 0, 1, 2, \ldots$. The union of these intervals becomes an a-random set in the state space $(L, X)$ when the lengths of the intervals are independently and identically distributed with the unit scale exponential density $\exp(-u)$ for $u \geq 0$.

Note that when you condition on a fixed $L = \lambda$ (by combining the Poisson DSM with a deterministic DSM with mass 1 on the set $\{(\lambda, x) : x \in \mathcal{N}\}$), the resulting DSM does indeed yield a precise $\mathcal{P}\text{ois}(\lambda)$ margin for $X$, since the number of unit exponential intervals that elapse in time $\lambda$ is described by a Poisson process with rate 1. Conditioning instead on a fixed count $k$ restricts the mass on the resulting DSM to the line corresponding to $X = k$. From Figure 5 we see that this places $L$ in the range $(V_k, V_{k+1})$ (letting $V_0 := 0$ for notational convenience). This characterization of inference about $L$ was first given in Almond (1989, 1995).

The left end $V_k$ of this a-random interval is defined by the sum of $k$ independent a-random unit-scale exponentials, and hence has a unit-scale gamma distribution with shape $k$. The length of the interval $(V_{k+1} - V_k)$ is independently exponentially distributed, and the $(V_k, V_{k+1})$ pair are jointly distributed as the $k$ and $k+1$ transition times of a unit-rate Poisson process. In words, observing that a unit-rate Poisson process has transitioned $k$ states in an unknown amount of time $\lambda$ provides evidence about $\lambda$ in the form of bounds: since in $\lambda$ time, $k$ exponentials elapsed, $\lambda > V_k$, where $V_k \sim \mathcal{G}\text{amma}(k)$, and since the next exponential has not yet elapsed, $\lambda < V_{k+1}$. The joint distribution of $(V_k, V_{k+1})$ is characterized by the formula

$$\mathbb{P}(V_k \leq u, V_{k+1} \geq v) = \frac{1}{k!} u^k \exp(-v) \qquad \forall v \geq u \geq 0.$$

In standard probability terms, this is a form of the bivariate cumulative distribution of the ends of the a-random interval $(V_k, V_{k+1})$. In DS terms, however, it is the commonality function $c(u, v)$ of the interval $(u, v)$ for the posterior DSM of $L$ given the observation $X = k$.

Note that $\lambda \in (V_k, V_{k+1})$ only when $V_k \leq \lambda$ and $V_{k+1} \geq \lambda$, so $\lambda \notin (V_k, V_{k+1})$ whenever $V_{k+1} < \lambda$ or $V_k > \lambda$. These are mutually exclusive events, so the plausibility $\text{Plaus}(\{\lambda\})$ of the singleton set $\{\lambda\}$, which is also the commonality of the trivial range $(\lambda, \lambda)$, is

$$\begin{aligned} \text{Plaus}(\{\lambda\}) = c(\lambda, \lambda) &= \mathbb{P}_{(V_k, V_{k+1})}(\lambda \in (V_k, V_{k+1})) \\ &= 1 - \mathbb{P}(\lambda \notin (V_k, V_{k+1})) \end{aligned}$$

$$= 1 - (F_{V_{k+1}}(\lambda) + (1 - F_{V_k}(\lambda)))$$
$$= F_{V_k}(\lambda) - F_{V_{k+1}}(\lambda).$$

4.2. *Join trees.* In the Banff challenge model in equation (1.1), the counts $y_i$ and $z_i$ are each from scaled Poison distributions with known, constant scale factors $t_i$ and $u_i$, respectively. There are at least three ways to extend the unscaled Poisson DSM as described above for use with scaled Poissons. Perhaps the simplest is to argue via representation that if an observation of $Y' = y'$ from a $\mathcal{P}\text{ois}(b)$ yields a posterior DSM on the $B$ margin with the commonality function described above, which bounds $b$ by $V_{y'}$ and $V_{y'+1}$, then an observation $Y = y$ from a $\mathcal{P}\text{ois}(tb)$ should yield a posterior DSM bounding $b$ by $\frac{V_y}{t}$ and $\frac{V_{y+1}}{t}$. Another approach would be to extend the definition of the Poisson DSM such that the auxiliary sequence $V_1, V_2, V_3, \ldots$ is separated by exponentials with scale $\frac{1}{t}$. In this section we introduce a third approach that uses join trees, which are a fundamental component of DS analysis (DSA) that greatly simplify the process of computing with DSMs.

If we define $\mathbf{L}_y := tb$, then (when the observed count is $Y = y$) the Poisson DSM as described in the previous section provides a posterior inference about $\mathbf{L}_y$ in the form of a DSM with commonality function

$$c(u, v) = \frac{1}{y!} u^y \exp(-v) \qquad \forall v \geq u \geq 0.$$

If we treat this as a DSM on the $\mathbf{L}_y$ margin of the larger SSM that also includes the $T$ and $B$ states, then we can extend this marginal DSM to the full SSM, and likewise extend the constant DSM $T = t$, combine them there and then project the resulting DSM to the $B$ margin to yield an inference about $b$.

The join tree theorem of DS analysis due to Shenoy and Shafer (1986) or Kong (1986) states that we may (as just described) first project the Poisson DSM over the $(\mathbf{L}_y, Y)$ SSM to the $\mathbf{L}_y$ margin, then extend from that margin up to the joint $(\mathbf{L}_y, T, B)$ SSM, combine there with the DSM about $T$ and then project to the $B$ margin (this process is called "propagation"). We do not need to extend all margins up to the full joint state space (which would include the $Y$, $\mathbf{L}_y$, $T$, and $B$ states) and combine there, since all relevant information on the $Y$ margin is contained in the projection of $(\mathbf{L}_y, Y)$ onto $\mathbf{L}_y$.

Figure 6 depicts the join tree for this example. A join tree is a hypergraph depicting nodes for each constituent margin of the SSM and hyperedges corresponding to each hint. In this example the hints are as follows:

- the Poisson DSM, relating $\mathbf{L}_y$ and $Y$,
- the observed count $Y = y$,
- the relationship given by the definition $\mathbf{L}_y := T \times B$, and

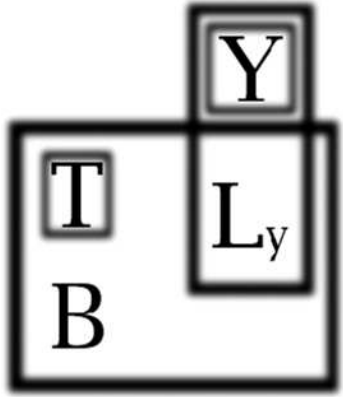

FIG. 6. *Join tree for a scaled Poisson DSM. A join tree is a hypergraph depicting nodes for each constituent margin of the SSM and hyperedges corresponding to each hint. Boxes depict the hints: $Y \sim \mathcal{P}\text{ois}(\mathbf{L}_y)$, where $\mathbf{L}_y := B \times T$, and $T$ and $Y$ are known constants.*

- the known scale $T = t$.

Note that if one hint refers to a group of margins $J$, and another refers to a group $K$, and $J \subseteq K$, then the two hints can be combined on the $K$ SSM. For example, the observed count $y$ and the Poisson DSM may be combined on the $(\mathbf{L}_y, Y)$ SSM. If we define $\mathcal{J}$ as those hyperedges remaining after all of these trivial combinations are performed, then the hyperedges in the *scheme* $\mathcal{J}$ are said to be nodes of a join tree if it is possible to define edges $\mathcal{E}$ (pairs of hyperedges in the scheme) such that:

- $(\mathcal{J}, \mathcal{E})$ is a tree, and
- if $J_i$ and $J_k$ are distinct vertices of the tree $(\mathcal{J}, \mathcal{E})$, then $J_i \cap J_k$ is contained in every vertex on the unique path from $J_i$ to $J_k$ in the tree.

The join tree for this example contains only two nodes, one formed from the Poisson DSM and one from the definition relating $\mathbf{L}_y$ to $B$ and $T$. The unique path between them includes no other nodes. If it did, then their intersection $\{\mathbf{L}_y\}$ would need to be contained in every node on that path.

The join tree theorem states that when such a tree exists, then DSA can be performed by propagating evidence from leaves of the join tree toward the node(s) containing the margin of interest (in this case, $B$). The full joint space of all states need never be constructed. In the present example this is a small convenience, but in larger problems (such as the Banff Challenge), the join tree theorem provides a significant reduction in computational complexity. For a more thorough treatment we refer the interested reader to any one of the many available tutorials on DS propagation [e.g., Almond (1988) or Kohlas and Monney (1994)].

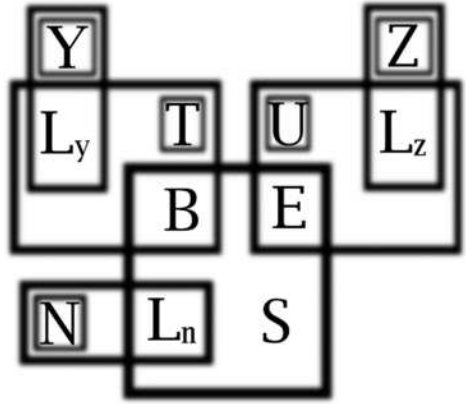

Fig. 7. *Join tree for the three-Poisson DSM. Boxes depict the hints given by the constituent Poisson DSMs. The overlap of the boxes depicts their shared components.*

4.3. *DS solution.* The join tree for the (single-channel) Banff Challenge is depicted in Figure 7. The components involving $\mathbf{L}_y$ and $\mathbf{L}_z$ are just as described in the previous section. The additional components relate the Poisson DSM on $(L_N, N)$ to the other components via the definition $L_N := ES + B$. An additional hint constrains $S$ to the nonnegative Reals (since the production rate of the Higgs particle must be $\geq 0$).

We will ultimately characterize our uncertainty about the quantity of interest $s$ by the distribution of an a-random variable $\mathbf{S}$. If $\mathrm{F}_{\mathbf{S}}(s) = \int_0^s \mathrm{f}_{\mathbf{S}}(x)\,dx$ is the cumulative distribution function of $\mathbf{S}$, then our goal is to find $s_{90}^*$ and $s_{99}^*$ such that $\mathrm{F}_{\mathbf{S}}(s_{90}^*) = 0.90$ and $\mathrm{F}_{\mathbf{S}}(s_{99}^*) = 0.99$. The function $\mathrm{f}_{\mathbf{S}}(\cdot)$ is the result of a *plausibility transformation* from the $S$ margin of the posterior DSM. In the $n$-channel case we get $\mathrm{f}_{\mathbf{S}}(x) \propto \prod_{i=1}^n r_i(x)$, where $r_i(x) = c_i(x, x)$ is the DS commonality of the singleton $\{x\}$ on the $S$ margin of the sub-DSM corresponding to channel $i$.

The DS commonality (for channel $i$) of the singleton $\{x\}$,

$$r_i(x) = (\mathrm{F}_{\mathbf{S}_l^i}(x) - \mathrm{F}_{\mathbf{S}_u^i}(x)),$$

is the difference between the CDFs of the a-random variables $\mathbf{S}_l^i$ and $\mathbf{S}_u^i$ for the lower end and upper end of the a-random interval. To complete the DS solution, we need the distribution functions $\mathrm{F}_{\mathbf{S}_l^i}(\cdot)$ and $\mathrm{F}_{\mathbf{S}_u^i}(\cdot)$. Equations for these functions are derived in the Appendix. A simplified result is provided here.

In the cases in which $n > 0$, $y > 0$, and $z > 0$, and ignoring the constraint that $s \geq 0$, the formulas are

$$\mathrm{F}_{\mathbf{S}_l^i}^*(x) = \mathbb{P}\bigg(\frac{\mathbf{N}_l^i - 1/t_i \mathbf{Y}_u^i}{\frac{1}{u_i}\mathbf{Z}_u^i} \leq x\bigg) \quad \text{and}$$

$$F^*_{\mathbf{S}^i_u}(x) = \mathbb{P}\bigg(\frac{\mathbf{N}^i_u - 1/t_i \mathbf{Y}^i_l}{1/u_i \mathbf{Z}^i_l} \leq x\bigg),$$

where $\mathbf{N}^i_l$, $\mathbf{Z}^i_l$, $\mathbf{Y}^i_l$ are the lower ends of the a-random intervals for $\mathbf{L}^i_n := (\varepsilon_i s + b_i)$, $\mathbf{L}^i_y := (t_i b_i)$, and $\mathbf{L}^i_z := (u_i \varepsilon_i)$, and $\mathbf{N}^i_u$, $\mathbf{Z}^i_u$, $\mathbf{Y}^i_u$ are the upper ends. The lower ends of these a-random intervals are distributed according to unit-scale independent gammas:

$$\mathbf{N}^i_l \sim \mathcal{G}\text{amma}(n_i),$$
$$\mathbf{Y}^i_l \sim \mathcal{G}\text{amma}(y_i) \quad \text{and}$$
$$\mathbf{Z}^i_l \sim \mathcal{G}\text{amma}(z_i),$$

respectively. The upper ends are also gamma distributed, such that the differences $\mathbf{N}^i_u - \mathbf{N}^i_l$, $\mathbf{Y}^i_u - \mathbf{Y}^i_l$, and $\mathbf{Z}^i_u - \mathbf{Z}^i_l$ are each independently exponentially distributed $\sim \mathcal{E}\text{xpo}(1)$.

**5. Discussion.** The principal benefit of the DS approach over the Bayesian is that, in DS analysis, information may be transmitted from margins of the space (in the present example, from the nuisance parameters $\varepsilon$ and $b$) to the joint space without requiring the use of a prior. The Bayesian calculus requires a prior because the map from the marginal space to the joint space must be one-to-one. In DS calculus it may be multivalued, so that the distribution of the smaller space over the larger space need not be specified.

The principal drawback, from a frequentist or classical Bayesian perspective, of the DS calculus is that the interpretation of a distribution over ranges (or more generally over sets of the state space) is awkward for those familiar with a single-valued framework. In this paper we map the distribution over ranges into a distribution over single values using the *plausibility transform*. This allows us to return a single value for use in comparing the method to other Bayesian and frequentist approaches.

One benefit of the DS approach is that, although priors are not required, prior information can be incorporated as easily as in Bayesian analyses. In the DS approach, however, as many or as few priors may be incorporated as there is prior information to incorporate. For instance, it may be the case that strong prior information exists for one of the nuisance parameters, but not for the other. This would easily be accommodated using the DS approach. If Bayesian priors are provided for all three of the parameters, this approach is exactly the same as the corresponding Bayesian approach. Priors may also be provided as DS models [typically referred to as "belief functions" in the literature, following Shafer (1976)], if the prior knowledge is better represented this way.

According to Banff Challenge organizer Joel Heinrich in Heinrich (2006c), "Subjective informative priors for the parameter of interest are very unpopular. We have no confidence in our own opinion, for example, of the mass of the Higgs particle, nor in anyone else's opinion. Therefore, even subjective priors are invariably uninformative for the parameter of interest. But priors for some nuisance parameters are, in some cases, both subjective and informative—a problem for frequentists." We argue that the DS approach is ideally suited for situations in which prior information exists for some, but not all, parameters of a model.

The DS approach described in this paper is conceptually straightforward, simple to implement, efficient to compute, and performs very well at the given tasks. Unlike pure Bayesian approaches, there is no need to specify a prior. If prior information is available, however, this approach can easily accommodate that information, unlike frequentist approaches.

It remains unclear whether the tasks of this challenge are representative of the actual scientific goal of placing confidence limits on the mass of the Higgs particle. The three-Poisson model for the background, efficiency, and combined (signal times efficiency plus background) counts may or may not be the best representation of the problem. These questions are perhaps best left to the physics and astronomy communities to debate.

What we have shown is that the DS approach, which heretofore has not been considered for the Poisson limits problem, is an approach at least on par with the more commonly considered techniques. DSA is a statistical framework that is not well understood by most statisticians, though it has steadily gained practitioners since its quiet inception in the middle of the last century. With the Bayesian-frequentist debate hinging primarily on the power and danger of incorporating prior information into an analysis, the Dempster–Shafer approach deserves consideration.

## APPENDIX: DERIVATION OF THE DS SOLUTION

Recall that our goal is to find $s_{90}^*$ and $s_{99}^*$ such that $F_{\mathbf{S}}(s_{90}^*) = 0.90$ and $F_{\mathbf{S}}(s_{99}^*) = 0.99$, where $F_{\mathbf{S}}(s) = \int_0^s f_{\mathbf{S}}(x)\,dx$.

The evidence given by the data corresponding to channel $i$, $(n_i, y_i, z_i, t_i, u_i)$, constrains the unknown $s$ to the a-random interval $(\mathbf{S}_l^i, \mathbf{S}_u^i)$. By characterizing the distribution functions of the a-random variables $\mathbf{S}_l^i$ and $\mathbf{S}_u^i$, we can calculate $r_i(x) = (F_{\mathbf{S}_l^i}(x) - F_{\mathbf{S}_u^i}(x))$, and from this, $f_{\mathbf{S}}(x) \propto \prod_{i=1}^n r_i(x)$.

We assume, for now, that $n_i > 0$, $y_i > 0$, and $z_i > 0$; the Special Cases section, in the Supplementary Materials, addresses the cases in which one or more of these counts is 0. By the Poisson DSM, we know that the distributions of the lower ends of the a-random intervals for $\mathbf{L}_n^i := (\varepsilon_i s + b_i)$, $\mathbf{L}_y^i := (t_i b_i)$, and $\mathbf{L}_z^i := (u_i \varepsilon_i)$ are independent unit-scale gammas:

$$\mathbf{N}_l^i \sim \mathcal{G}\mathrm{amma}(n_i),$$
$$\mathbf{Y}_l^i \sim \mathcal{G}\mathrm{amma}(y_i) \quad \text{and}$$
$$\mathbf{Z}_l^i \sim \mathcal{G}\mathrm{amma}(z_i),$$

respectively, and that the upper ends are also gamma distributed, such that the differences $\mathbf{N}_u^i - \mathbf{N}_l^i$, $\mathbf{Y}_u^i - \mathbf{Y}_l^i$, and $\mathbf{Z}_u^i - \mathbf{Z}_l^i$ are each independently exponentially distributed $\sim \mathcal{E}\text{xpo}(1)$. From this, we get the constraints that

$$\mathbf{N}_l^i \leq (\varepsilon_i s + b_i) \leq \mathbf{N}_u^i,$$

$$\frac{1}{t_i}\mathbf{Y}_l^i \leq b_i \leq \frac{1}{t_i}\mathbf{Y}_u^i \quad \text{and}$$

$$\frac{1}{u_i}\mathbf{Z}_l^i \leq \varepsilon_i \leq \frac{1}{u_i}\mathbf{Z}_u^i.$$

From these and from the additional constraint that $s \geq 0$, we see that

$$\mathbf{S}_l^i = \frac{\max(0, \mathbf{N}_l^i - 1/t\mathbf{Y}_u^i)}{1/u\mathbf{Z}_u^i} \quad \text{and}$$

$$\mathbf{S}_u^i = \frac{\mathbf{N}_u^i - 1/t\mathbf{Y}_l^i}{1/u\mathbf{Z}_l^i}$$

in the equation $\mathbf{S}_l^i \leq s \leq \mathbf{S}_u^i$.

Thus, if we ignore (momentarily) the constraint that $s \geq 0$, we may characterize the CDFs of $\mathbf{S}_l^i$ and $\mathbf{S}_u^i$ as

$$F^*_{\mathbf{S}_l^i}(x) = \mathbb{P}\left(\frac{\mathbf{N}_l^i - 1/t_i\mathbf{Y}_u^i}{1/u_i\mathbf{Z}_u^i} \leq x\right) \quad \text{and}$$

$$F^*_{\mathbf{S}_u^i}(x) = \mathbb{P}\left(\frac{\mathbf{N}_u^i - 1/t_i\mathbf{Y}_l^i}{1/u_i\mathbf{Z}_l^i} \leq x\right).$$

Rearranging, we may write this as

$$F^*_{\mathbf{S}_l^i}(x) = \mathbb{P}\left(\mathbf{N}_l^i \leq \frac{1}{t_i}\mathbf{Y}_u^i + \frac{x}{u_i}\mathbf{Z}_u^i\right) \quad \text{and}$$

$$F^*_{\mathbf{S}_u^i}(x) = \mathbb{P}\left(\mathbf{N}_u^i \leq \frac{1}{t_i}\mathbf{Y}_l^i + \frac{x}{u_i}\mathbf{Z}_l^i\right).$$

We are ultimately interested in the normalized quantities

$$F_{\mathbf{S}_l^i}(x) = \frac{F^*_{\mathbf{S}_l^i}(x) - \mathbb{P}(\mathbf{S}_u^i < 0)}{1 - \mathbb{P}(\mathbf{S}_u^i < 0)} \quad \text{and}$$

$$F_{\mathbf{S}_u^i}(x) = \frac{F^*_{\mathbf{S}_u^i}(x) - \mathbb{P}(\mathbf{S}_u^i < 0)}{1 - \mathbb{P}(\mathbf{S}_u^i < 0)},$$

where we condition on the upper end of the interval, $\mathbf{S}_u^i$, being nonnegative. Since this condition is met whenever $\mathbf{N}_u^i \geq \frac{1}{t_i}\mathbf{Y}_l^i$, we have

$$F_{\mathbf{S}_l^i}(x) = \frac{\mathbb{P}(\mathbf{N}_l^i \leq 1/t_i\mathbf{Y}_u^i + x/u_i\mathbf{Z}_u^i) - \mathbb{P}(\mathbf{N}_u^i < 1/t_i\mathbf{Y}_l^i)}{1 - \mathbb{P}(\mathbf{N}_u^i < 1/t_i\mathbf{Y}_l^i)} \quad \text{and}$$

(A.1)
$$F_{\mathbf{S}_u^i}(x) = \frac{\mathbb{P}(\mathbf{N}_u^i \leq 1/t_i \mathbf{Y}_l^i + x/u_i \mathbf{Z}_l^i) - \mathbb{P}(\mathbf{N}_u^i < 1/t_i \mathbf{Y}_l^i)}{1 - \mathbb{P}(\mathbf{N}_u^i < 1/t_i \mathbf{Y}_l^i)}.$$

As we show in more detail in the Supplementary Materials, these may be expressed in terms of the $\mathcal{B}$eta CDF as

$$F_{\mathbf{S}_l^i}(x) = 1 - \bigg( p\mathcal{B}(\alpha_i, z_i + 1, n_i)$$
$$- \int_0^{\alpha_i} p\mathcal{B}\bigg(\frac{1}{1 + t_i(1 - \gamma/\alpha_i)},$$
$$z_i + 1 + n_i, y_i + 1\bigg) d\mathcal{B}(\gamma, z_i + 1, n_i) \, d\gamma \bigg)$$
$$\times \bigg( p\mathcal{B}\bigg(\frac{t_i}{t_i + 1}, y_i, n_i + 1\bigg)\bigg)^{-1} \quad \text{and}$$

(A.2)
$$F_{\mathbf{S}_u^i}(x) = 1 - \bigg( p\mathcal{B}(\alpha_i, z_i, n_i + 1)$$
$$- \int_0^{\alpha_i} p\mathcal{B}\bigg(\frac{1}{1 + t_i(1 - \gamma/\alpha_i)},$$
$$z_i + n_i + 1, y_i\bigg) d\mathcal{B}(\gamma, z_i, n_i + 1) \, d\gamma \bigg)$$
$$\times \bigg( p\mathcal{B}\bigg(\frac{t_i}{t_i + 1}, y_i, n_i + 1\bigg)\bigg)^{-1},$$

where $d\mathcal{B}(\cdot, \alpha, \beta)$ is the PDF of a $\mathcal{B}$eta distribution with parameters $\alpha$ and $\beta$, $p\mathcal{B}(\cdot, \alpha, \beta)$ is its CDF, and $\alpha_i := \frac{u_i}{u_i + x}$.

**Acknowledgments.** The authors would like to acknowledge Mathis Thoma, Joel Heinrich, and the members of the California/Harvard Astrostatistics Collaboration (CHASC) for their inspiration and helpful feedback during the course of this project.

## SUPPLEMENTARY MATERIAL

**Supplement A: A complete derivation of the DS solution to the Banff challenge** (DOI: 10.1214/08-AOAS223SUPP; .pdf). We provide a complete derivation of the Dempster–Shafer solution, including detail and special cases not covered in the Appendix.

P. T. EDLEFSEN
A. P. DEMPSTER
STATISTICS DEPARTMENT
HARVARD UNIVERSITY
SCIENCE CENTER
1 OXFORD STREET
CAMBRIDGE, MASSACHUSETTS 02138-2901
USA
E-MAIL: edlefsen@fas.harvard.edu
       dempster@fas.harvard.edu

C. LIU
DEPARTMENT OF STATISTICS
PURDUE UNIVERSITY
250 N. UNIVERSITY STREET
WEST LAFAYETTE, INDIANA 47907-2066
USA
E-MAIL: chuanhai@stat.purdue.edu